\documentclass[multphys,vecphys]{svmult}

\usepackage{graphicx}    

\begin{document}

\title*{Radio Monitoring of Supernova 2001ig:\\ The First Year}
\titlerunning{Radio Monitoring of SN~2001ig}
\author{Stuart D. Ryder\inst{1}, Elaine Sadler\inst{2},
Ravi Subrahmanyan\inst{3}, Kurt W. Weiler\inst{4}, Nino Panagia\inst{5}\and
Christopher Stockdale\inst{4,6}}
\authorrunning{S. D. Ryder et al.}
\institute{Anglo-Australian Observatory, P.O. Box 296, Epping, NSW 1710,
Australia
\texttt{sdr@aaoepp.aao.gov.au}
\and School of Physics, University of Sydney, NSW 2006, Australia
\texttt{ems@physics.usyd.edu.au}
\and Australia Telescope National Facility, CSIRO, Locked Bag 194, Narrabri,
NSW 2390, Australia
\texttt{Ravi.Subrahmanyan@csiro.au}
\and Naval Research Laboratory, Code 7213, Washington, DC 20375-5320, U.S.A.
\texttt{Kurt.Weiler@nrl.navy.mil}
\and ESA/Space Telescope Science Institute, 3700 San Martin Drive, Baltimore,
MD~21218, U.S.A.
\texttt{panagia@stsci.edu}
\and Physics Dept., Maquette University, P.O. Box 1881, Milwaukee, WI~53201,
U.S.A.
\texttt{christopher.stockdale@mu.edu}
}
%
%
\maketitle

\abstract{
Supernova 2001ig in NGC~7424 has been observed with the Australia
Telescope Compact Array at $\sim$2~week intervals since its discovery,
making this the best-studied Type IIb radio supernova since SN~1993J. We
present radio light curves for frequencies from 1.4 to 20\,GHz, and
preliminary attempts to model the observed behaviour. Since peaking in radio
luminosity at 8.6 and 4.8\,GHz some 1-2 months after the explosion, SN~2001ig
has on at least two occasions deviated significantly from a smooth decline,
indicative of interaction with a dense circumstellar medium and possibly of
periodic progenitor mass-loss.}

\section{Introduction}
\label{s:intro}

On the evening of Dec 10 2001, the Rev. Robert Evans found his 39th
supernova from his home in the Blue Mountains, west of Sydney
\cite{evans01}. SN~2001ig lies in the outskirts of the SAB(rs)cd
galaxy NGC~7424, at a distance of 11.5\,Mpc \cite{tu88}. Early optical
spectroscopy with the 6.5\,m Baade Telescope by Matheson \& Jha
\cite{mj01} highlighted several similarities between SN~2001ig and the
Type~IIb SN~1987K \cite{fil88}. In the months following, the spectral
evolution of SN~2001ig began to resemble more and more that of the
``prototypical'' Type~IIb SN~1993J, as the H recombination lines faded
\cite{cp01,clo02}, and eventually disappeared \cite{fc02}.

During a Director's Discretionary Time observation, SN~2001ig was
detected by the ACIS-S instrument on board the \emph{Chandra}
X-ray Observatory on 2002 May~22 UT. A total of 30~counts was recorded
in 23400 sec of integration, corresponding to a 0.2-10.0~keV luminosity
$\sim10^{38}$~erg~s$^{-1}$ \cite{sr02}.

Since its commissioning, the Australia Telescope Compact Array
(ATCA)\footnote{The Australia Telescope is funded by the Commonwealth
of Australia for operation as a National Facility managed by CSIRO.}
has played a leading role in the monitoring of several supernovae at
radio wavelengths, most notably SN~1987A \cite{lss92}, SN~1978K
\cite{sdr93,ems99}, and SN~1998bw/GRB980425 \cite{wkf99}. As SN~2001ig
was too far south to allow effective monitoring with the Very Large
Array (VLA) in its most compact (D) configuration at the time, we commenced
observations with the ATCA within a week of its discovery, and have
been following it on a regular basis since then.


\section{Radio Light Curves}
\label{s:rlc}

\begin{figure}
\centering
\includegraphics[angle=270,width=11cm]{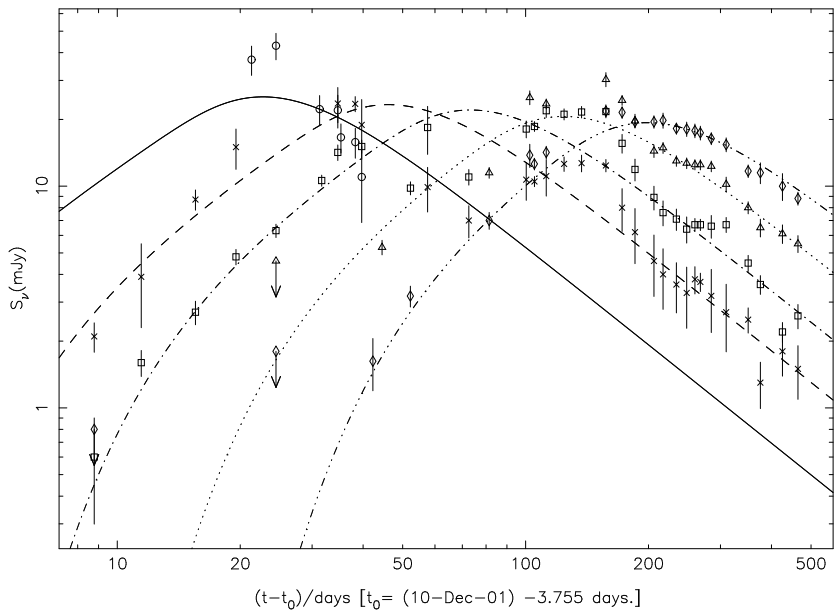}
\caption{Radio ``light curves'' for SN~2001ig at frequencies of
22.5/18.8\,GHz (\emph{circles, solid line}); 8.6\,GHz (\emph{crosses,
dashed line}); 4.8\,GHz (\emph{squares, dash--dotted line}); 2.4\,GHz
(\emph{triangles, dotted line}); and 1.4\,GHz (\emph{diamonds, dash--triple
dotted line}). The curves are model fits to the data, as described in the
text.}
\label{f:kw}
\end{figure}

Our first ATCA observations of SN~2001ig over 6~hours on 2001 Dec~15
UT yielded a positive detection at 8.64\,GHz, and a marginal detection
at 4.79\,GHz \cite{sr01}. On 2001 Dec~31 UT, we detected SN~2001ig
using a prototype 18.8\,GHz receiver system on just three ATCA
antennas. Observations have been carried out with a 128\,MHz
bandwidth, centered on the primary frequencies of 8.64, 4.79, 2.50,
and 1.38\,GHz, supplemented by some higher frequency data from both
the ATCA and the VLA. The primary flux calibrator for the ATCA data is
PKS~B1934-638, while the source PKS~B2310-417 serves as the secondary
gain and phase calibrator. The primary beam around SN~2001ig
happens to include an adjacent background source just
$20^{\prime\prime}$ away, which has also proved useful for gain
calibration when phase stability was poor, or hour-angle coverage
limited. Figure~\ref{f:kw} presents our entire dataset up to 2003
March~16.

The radio ``light curve'' of a supernova typically proceeds through
three phases -- a rapid turn-on, with a spectral index which is
inverted ($\alpha = 2$ or steeper, where $S_{\nu} \propto \nu ^
{\alpha}$) due to absorption along the line-of-sight; followed by a
peak in the flux density, firstly at the higher frequencies; then a
more gradual decline in the optically-thin phase, with a non-thermal
spectral index. By the end of February 2002, SN~2000ig had already
peaked at frequencies of 8.64\,GHz and 4.79\,GHz. However, in early
March, the fluxes at these two frequencies jumped by a factor of 2,
and remained almost constant for the next two months, before resuming
their decline. In August 2002, there was another short but significant
pause in the decline. Similar, but less-pronounced deviations are
also apparent in the data at 2.50 and 1.38\,GHz.

Superimposed on Fig.~\ref{f:kw} are model fits to the multi-frequency
dataset, based on the ``minishell'' model of \cite{che82}, as
parameterised by \cite{kw02}:

\begin{eqnarray*} 
S {\rm (mJy)} & = & K_1 {\left({\nu} \over {\rm 5~GHz}\right)^{\alpha}}
{\left({t - t_0} \over {\rm 1~day}\right)^{\beta}} e^{-{\tau}}
{\left({1 -e^{-{\tau^{\prime}}}} \over {\tau^{\prime}} \right)} \; , \\
{\rm where~~} \tau & = & K_2 {\left({\nu} \over {\rm 5\,GHz}\right)^{-2.1}}
{\left({t - t_0} \over {\rm 1\,day}\right)^{\delta}} \\
{\rm and~~} \tau^{\prime} & = & K_3 {\left({\nu} \over {\rm 5\,GHz}\right)^
{-2.1}}
{\left({t - t_0} \over {\rm 1\,day}\right)^{\delta^{\prime}}}
\end{eqnarray*}

\noindent
Here $K_1$ is the flux density, $K_2$ and $K_3$ the attenuation by a
homogeneous, and a clumpy absorbing medium respectively, at a
frequency of 5~GHz one day after the explosion date $t_0$; $\alpha$ is
the spectral index; $\beta$ the rate of decline in the optically-thin
phase; while $\delta$ and $\delta^{\prime}$ describe the time
dependence of the optical depths in the homogeneous, and clumpy
circumstellar medium (CSM) respectively. By constraining the model
fits using only the data leading up to the high-frequency turnover, as
well as the late-time decay, we can at least compare the global
characteristics of this event with those of the best-studied Type~IIb
supernova, SN~1993J (Table~\ref{t:fits}, \cite{svd03}). We note that
despite the temporary ``boosts'' in the radio flux, the overall rate
of decline $\beta$ in SN~2001ig is still much faster than SN~1993J.
Furthermore, while the model correctly predicted a 5\,GHz peak
luminosity twice that of SN~1993J, it was not actually attained until
after day~100.

\begin{table}
\centering
\caption{Comparison of radio light curve model parameters.\label{t:fits}}
\begin{tabular}{ccc}
\hline\noalign{\smallskip}
Parameter    &     ~~~SN 2001ig~~~      &      ~~~SN 1993J~~~      \\
\noalign{\smallskip}\hline\noalign{\smallskip}
$K_1$ (mJy)  & $2.47\times10^{4}$ & $1.36\times10^{4}$ \\
$\alpha$     &     $-1.07$        &     $-1.05$        \\
$\beta$      &     $-1.50$        &     $-0.88$        \\
$K_2$        & $1.13\times10^{2}$ & $9.14\times10^{2}$ \\
$\delta$     &     $-1.94$        &     $-1.88$        \\
$K_3$        & $1.26\times10^{5}$ & $8.33\times10^{4}$ \\
$\delta^{\prime}$ & $-2.69$       &     $-2.26$        \\
$L_{\rm 5\ GHz \ peak}$ (erg s$^{-1}$ Hz$^{-1}$) &
            $3.5\times10^{27}$ & $1.4\times10^{27}$    \\
\noalign{\smallskip}\hline
\end{tabular}
\end{table}

\section{Periodic mass-loss}
\label{s:disc}

\begin{figure}
\centering
\includegraphics[angle=270,width=11cm]{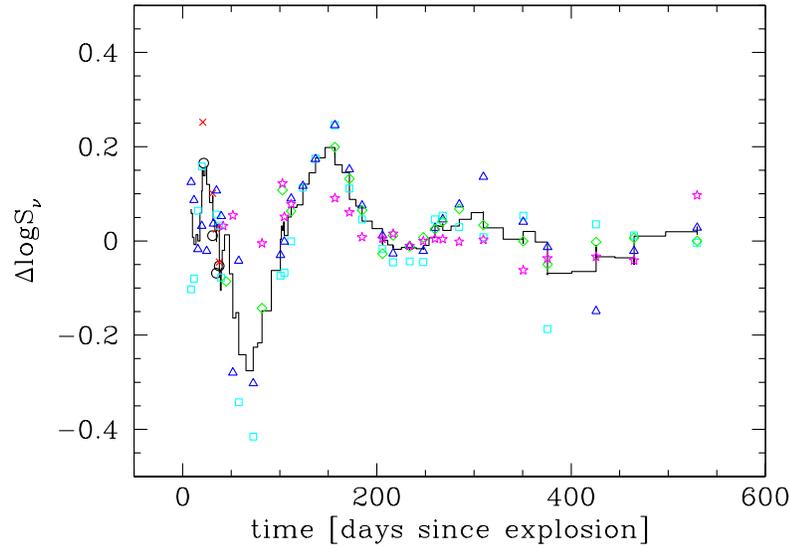}
\caption{Deviations of the observed flux of SN~2001ig about the
best-fit model. The symbols are the same as in Fig.~\protect{\ref{f:kw}}.
The solid line is a 4-point boxcar average of the mean deviation over all
frequencies at each epoch.}
\label{f:devs}
\end{figure}

Bumps and dips in the radio light curve can arise from modulations in
either the optical depth, or the CSM density structure (both of which
are coupled to some extent) via enhanced mass-loss. Throughout all
these events, the spectral index $\alpha$ is relatively unaffected,
which leads us to favour the latter mechanism. A constant $\alpha$
implies that the percentage change in flux is the same at all
frequencies, and this is illustrated in Fig.~\ref{f:devs} on which
is plotted the fractional deviation of the observed flux density
from the best-fit model curves in Fig.~\ref{f:kw}, as a linear function
of time. Figure~\ref{f:devs} also highlights an apparent periodicity in
CSM density enhancements, with the shock wave reaching the peak of the
first at $t\sim$150~days, a second peak near 300~days, and hints of a
third peak between 400 and 500~days. For an ejecta expansion velocity
of 15000~km~s$^{-1}$ and a stellar wind velocity of 10~km~s$^{-1}$,
this would imply a series of shells $\sim$0.006\,pc apart, ejected
every 600\,years or so. This is much longer than standard stellar
pulsation timescales, but is not inconsistent with the period between
thermal pulses (C/He shell flashes) in 5-10~M$_{\odot}$ AGB stars
\cite{ir83}.

Only SN~1979C has shown such regular structure in its radio light
curve \cite{kw92}, though these variations eventually ceased
\cite{mm00}. The proposed explanation was modulation of the red
supergiant progenitor wind due to eccentric orbital motion with a
4000~year period about a massive companion. Acceleration of the
progenitor near periastron would result in wind density enhancements
superimposed on a pinwheel-like CSM structure, which can account for
the periodicity in the radio emission \cite{sp96}. However, the
variations will only be pronounced if the orbital plane is viewed from
close to edge-on, which would then naturally account for why so few
supernovae display such regular variations in their radio light
curves.

Direct evidence for the existence of binary-generated spiral shocks comes
from near-infrared aperture-masking interferometry on the Keck~I telescope
of the Wolf-Rayet stars WR~104 \cite{tmd99} and WR~98a \cite{mtd99}. Both
sources show pinwheel-shaped nebulae, which are attributed to dust
forming where the stellar winds of the WR star and an OB-type companion
collide, then splaying outward. These sources may represent local,
compact analogs of the CSM structure that is currently being swept
up by the expanding blast wave from SN~2001ig.

\vskip 3mm
\emph{Acknowledgements:} We are grateful to the staff of the ATNF Paul Wild
Observatory for their assistance with the ATCA observations.




\end{document}